\documentstyle[12pt,psfig,epsfig]{article}
\input epsf

\begin{document}
\mbox{ }
\vspace{0.10in}\mbox{ }\\
\begin{center}
{\bf Time Dependence of Chemical Freeze-out in Relativistic Heavy Ion Collisions}\\
\bigskip
\bigskip
R. Bellwied$^1$, H. Caines$^2$, and T.J. Humanic$^2$\\

\medskip

{\em $^1$ Wayne State University, Physics Department, Detroit, Michigan}

\medskip

{\em $^2$ The Ohio State University, Physics Department, Columbus, Ohio}

\bigskip

(23-Mar-00, submitted to Phys.Rev.C)\\
\end{center}

\bigskip
\medskip
\noindent{\bf Introduction}
\medskip

Recent model calculations, which were successfully applied to
CERN SPS data, implied that most measured particle distributions
in relativistic heavy ion collisions
can be described with statistical models \cite{beca}. The underlying 
kinematics
requires at least thermal equilibration at freezeout. All kinetic
spectra, like the transverse momentum and the rapidity distributions,
can be explained with a single freezeout temperature, if one assumes
a certain expansion velocity after hadronization. Attempts to describe
not only the kinetic spectra but also the particle abundances and
particle ratios with a common temperature were successful, but led
to a different temperature. Whereas the slopes of all momentum spectra
yield a T around 130 MeV after inclusion of the expansion velocity,
the particle ratios lead to a T of about 180 MeV for a common chemical
potential. Two questions arise: is the chemical freezeout decoupled
from the thermal freezeout, and is the system chemically equilibrated
at chemical freezeout ? In a systematic study of measured particle ratios and
spectra in the framework of a thermal model, Heinz has shown that for all CERN 
and AGS data the chemical freezeout seems to be decoupled from the thermal 
freeze-out \cite{heinz}.

Rafelksi has recently postulated that, on the basis of strange particle
ratios measured in Pb+Pb at CERN, one has to assume a phase transition from 
QGP to Hadron Gas, and that at the time of hadronization the system is not in 
chemical equilibrium \cite{rafelski1}. To describe the kinetic spectra and
particle yields this model requires that the chemical composition of the
fireball remains unchanged between hadronization and thermal freeze-out.
That means all interactions during this time interval have to be elastic.
This model is in agreement with the fact that all systems from 
e$^{+}$e$^{-}$-collisions up to AA-collisions seem to yield the same chemical 
freezeout temperature \cite{beca}. Thus, the shapes of the particle spectra 
might change due to rescattering, but the particle abundances and ratios 
remain constant from hadronization on, a theory which is commonly referred 
to as 'sudden hadronization'. This process explicitly prohibits hadro-chemical 
equilibration through rescattering. Only elastic processes can be employed to 
explain the difference between the temperature calculated from particle 
ratios and the temperature calculated from kinetic particle spectra. 
Thus, the ratios 
effectively reflect the temperature at hadronization, whereas
the kinetic freezeout temperature describes the actual freezeout (last 
elastic final state interaction). The fact that the hadronization temperature 
is about 180 MeV, close to the critical temperature, can be viewed as an 
indicator that the system actually crossed a phase transition. However, the 
fact 
that all inelastic scattering processes cease at hadronization time is a very 
strong constraint. Hadronic population ratios are, in this theory, the result 
of the hadronization mechanism and not caused by interactions
during the hadronic rescattering phase. The main argument employed is that the 
relaxation times of all relevant hadronic channels is well above $\tau$=3 fm/c 
at which time the temperature of the system has dropped below T=185 MeV which 
leads to a small probability of chemical equilibration after hadronization. 
Also based on results of HBT measurements, Stock argues
that the time between hadronization and thermal freezeout might
simply be too short to develop a significant inelastic cross section
contribution \cite{stock}. 
The required time interval for hadronic processes to adjust the strangeness 
content is long and the chemical rates are small since the production of 
pairs of strange hadrons
carries a large energy penalty factor. In Stock's scenario it takes upward 
of 3 fm/c to equilibrate strangeness in the hadronic phase, a time span that,
he postulates, is not available owing to the rapid expansion prevailing at 
hadronization time. 

Surprisingly, a partonic cascade approach, which is very different from the 
model described above, yields results similar to a thermal sudden
hadronization theory. Geiger and Ellis \cite{geiger} have 
shown that at large incident energies the hadronization process can determine 
the final particle ratios based on its combined non-perturbative mechanisms
and it can even lead to direct chemical equilibrium.
The final multihadronic state materializes into maximal entropy (equilibrium) 
straight out of the partonic phase. Although certain remnants of the initial 
particle structure functions remain, the hadron yield near midrapidity stems 
mostly from the initial parton cascade processes. In this theory, the partonic
phase exhibits $\epsilon$ $\geq$ 2 Gev/fm$^{3}$ until about 2 fm/c. 
Hadronization occurs after a formation time of about 1 fm/c and a mixed phase
of hadrons and partons ends at $\tau$ $\approx$ 20 fm/c, when $\epsilon$ is 
less than 0.2 GeV/fm$^{3}$. Although the concept of sudden hadronization
is lost in this theory, due to the required long mixed phase between partons
and hadrons, the results are comparable to a thermal model without inelastic
final state interactions. 
While the parton models require a mixed phase for the proper kinematic
expansion of the system, they do not require any final state interactions
to describe the measured particle abundances.
Kapusta and Mekjian \cite{kapusta} derived estimates 
for the dynamical equilibration (relaxation) times of quark flavors in 
a quark gluon gas and deduced predictions for several equilibrium abundance 
ratios. These ratios are in good agreement with the data, which led
Stock \cite{stock} to postulate that the equilibrated ratios formed in 
pre-hadronization reactions do not change during the final state interactions.

\medskip

The main motivation for this paper is the fact that most dynamic
transport codes, which in the past were successfully applied to
describe data from CERN and the AGS, behave very differently from either the 
parton cascade or the thermal sudden hadronization approach.
These models employ measured inelastic scattering cross sections for different 
particle species to describe the hadronic transport from
hadronization to kinetic freezeout. Many of the relevant particle species 
are susceptible to number changing interactions, in particular inelastic meson 
interactions lead to sizable contributions
to the hyperon production cross section. Both, strangeness creation
interactions (e.g. $\pi$N $\rightarrow$ K$\Lambda$) and strangeness exchange
interactions (e.g. KN $\rightarrow$ $\pi\Lambda$) can be exothermic and should
affect strange particle production.

Pratt and Haglin have recently pointed out that number changing cross 
sections after hadronization seem to be relevant for the description 
of the pion abundance at CERN \cite{pratt}. They have shown that hadrons 
interact several times before the freezeout temperatures are reached 
if one assumes that binary modeling starts at T=160 MeV in a chemically and 
thermally equilibrated system. It is interesting to note that the chemical 
equilibration times for strange and non strange quarks shown by Stock and 
Pratt/Haglin are very comparable. An independent study by Humanic, using an 
early version of the transport code which is the basis of this paper, shows 
that pions interact on average ten times between hadronization and freezeout 
\cite{humanic2}. The time from hadronization to freezeout is quite long 
(about 15 fm/c) in particular for the lightest mesons. If these transport 
models are correct then we should be able to simulate a quantitative
dynamic evolution of all particle abundances through the rescattering phase. 

\bigskip
\medskip
\noindent{\bf Hadronic particle ratio simulations}
\medskip

In the following we attempt to prove that in particular the abundance of the 
singly strange baryon, the $\Lambda$, is affected by a series of inelastic 
rescattering processes by the time thermal freeze-out is accomplished.
We have employed the dynamic transport code described
in Ref.\cite{humanic2} to model the initial state of the system at
hadronization and the subsequent hadronic rescattering to freeze out. 
This code has been shown to well represent the features and dynamic
dependences of the transverse mass spectra and HBT observables for
CERN SPS Pb+Pb data \cite{humanic2}. 
In addition to the already implemented cross sections for pions, kaons,
nucleons and their associated resonances, we have
augmented the code to include elastic and inelastic rescattering of
$\Lambda$ baryons. The initial conditions are described by a common 
temperature and spatial extension. All particles hadronize at a proper
time of 1 fm/c. A Bjorken-type geometry was used to simulate the dynamic
evolution of the fireball from hadronization time. No initial radial flow is
needed for the calculations to agree with data \cite{humanic2}.

\medskip

Fig.1 shows the effect of rescattering on all four relevant particle
abundances ($\pi$, K, nucleons, and $\Lambda$).

\medskip

\begin{figure}[hbtl]
\hspace{1.in}
\psfig{figure=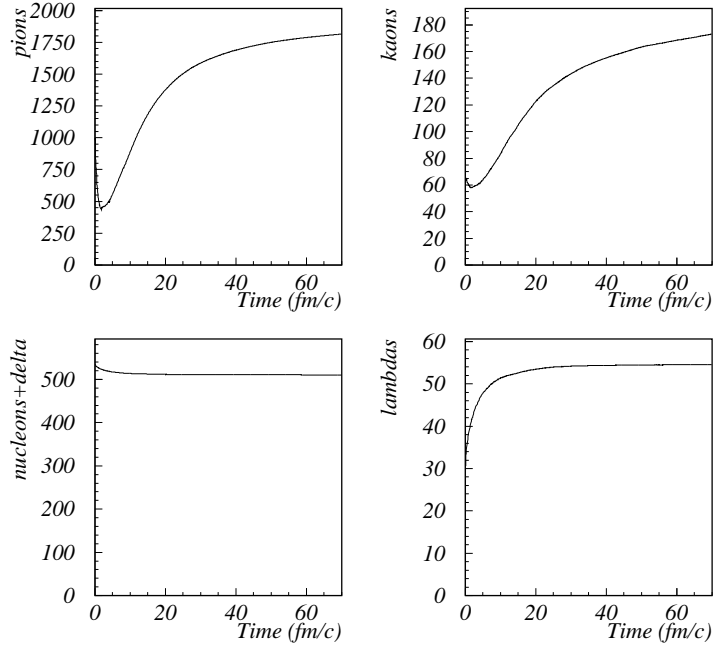,height=4.in}
\caption{Particle abundances as a function of time between hadronization
and kinematic freeze-out}
\label{fig:fig.1}
\end{figure}

\medskip

In the case of the nucleons, the nucleon resonance production and subsequent
decay leads to no change in the number of nucleons and is thus considered
an elastic process, whereas in the pion case, generation from
resonance decay is considered a number changing process (e.g. 
N$\pi$ $\rightarrow$ $\Delta\pi$ $\rightarrow$ N$\pi\pi$). The same is true
for vector meson production ($\rho$,$\phi$,$\omega$) and decay into pions.
Fig.1 shows that
the nucleon abundance is close to constant, but the meson and strange
baryon abundances display a steady increase.
One of the large contributing elastic reactions is 
$\pi\pi$ $\rightarrow$ $\rho$ $\rightarrow$ $\pi\pi$ 
which is fast and keeps the number of $\rho$'s in equilibrium but does 
not change the pion number. Reactions that change the overall pion number 
like $\pi\pi$ $\rightarrow$ $\eta^{'}$ $\rightarrow$ $\pi\pi\pi$ are generally 
slower and thus do not have a large effect on the chemical state formed at 
hadronization. In addition reactions that conserve the net number of 
strange quarks occur 
rapidly, e.g. KN $\rightarrow$ $\Lambda\pi$, but reactions that change the 
number of strange quarks, e.g. $\pi$N $\rightarrow$ K$\Lambda$, are slow as 
they  require a strange and an antistrange hadron to either interact or to 
be produced jointly. 

\medskip

Fig.2 shows the relative contributions of all relevant reaction channels
to the $\Lambda$ abundance as a function of time for the Pb+Pb and
the S+S systems at the SPS. In the Pb case the direct $\Lambda$
production at hadronization yields only around 50\% ($\approx$
30 $\Lambda$'s) of the total yield at kinetic freezeout. It is apparent
that $\Lambda$ generation processes dominate the $\Lambda$ inelastic 
re-scattering, which effectively leads to a factor two enhancement 
in $\Lambda$ production well after hadronization. In the smaller S+S
system the rescattering contributes only about 10\% of the total $\Lambda$
yield. 

\medskip

\begin{figure}[hbtl]
\begin{center}
\mbox{
\epsfig{file=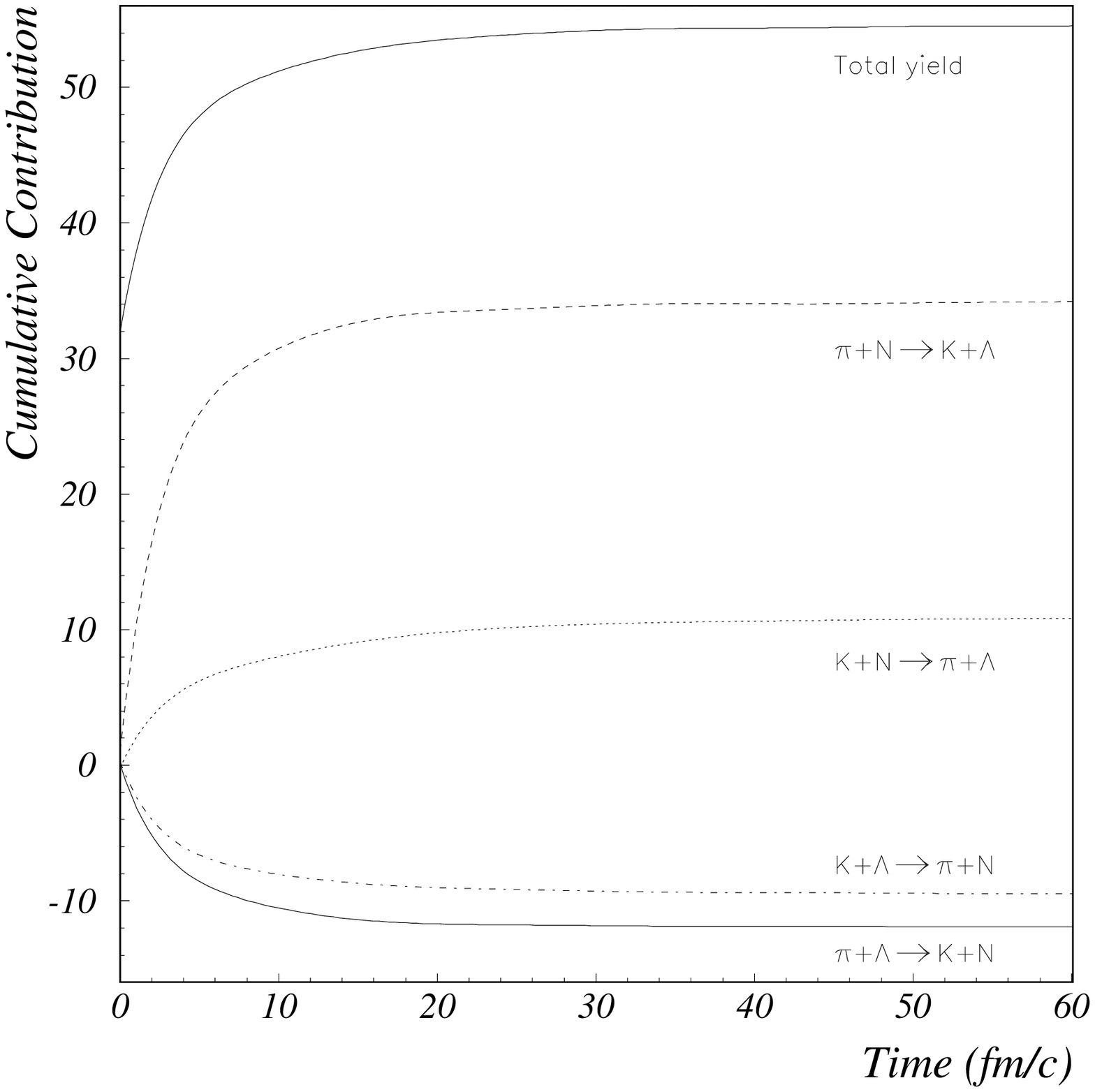,bbllx=50pt,bblly=155pt,bburx=520pt,bbury=645pt,width=6.5cm}
}
\mbox{
\epsfig{file=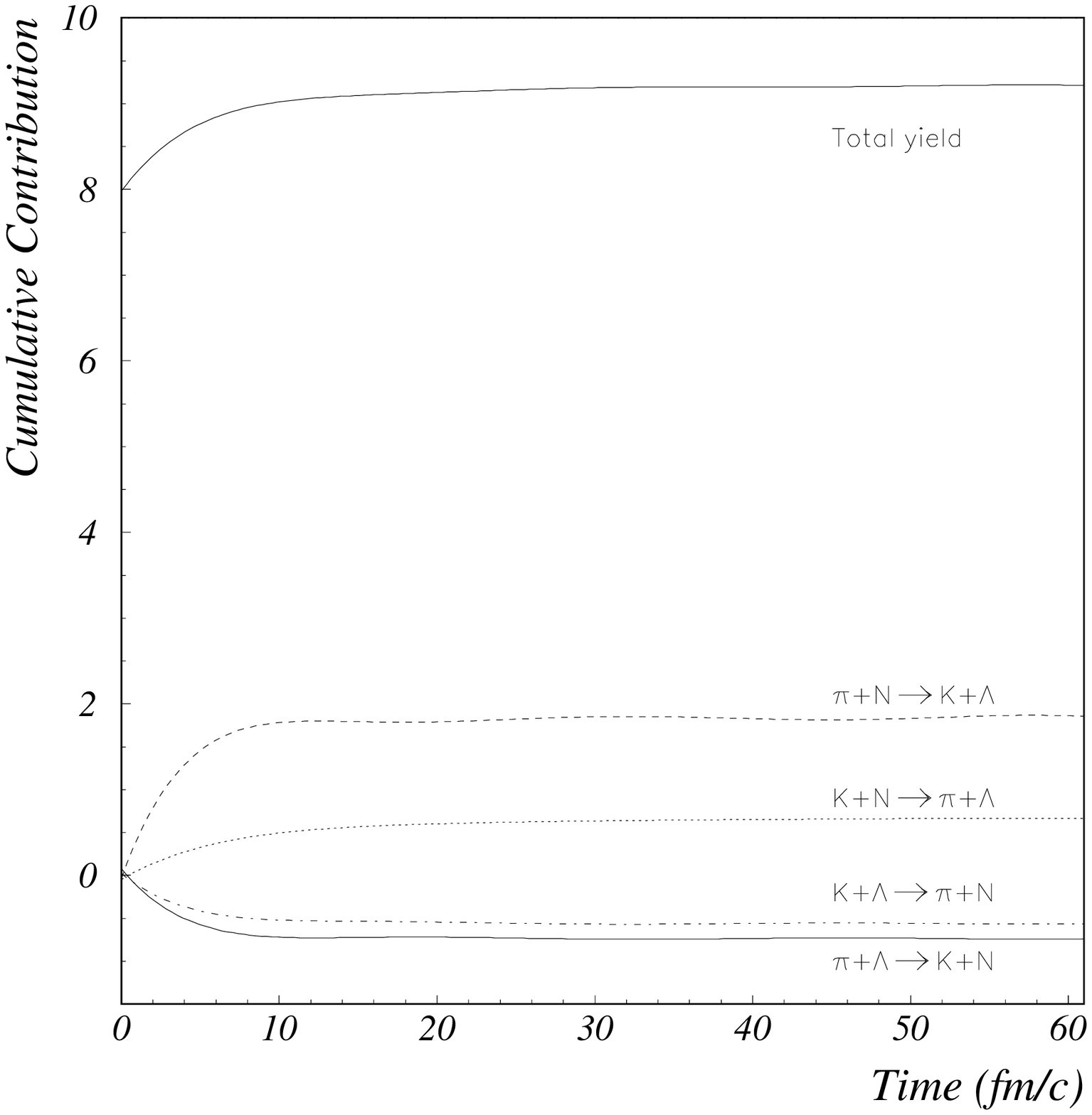,bbllx=50pt,bblly=155pt,bburx=520pt,bbury=645pt,width=6.5cm}
}
\end{center}
\caption{Contributing channels to the $\Lambda$ cross section in a.) Pb+Pb
and b.) S+S}
\label{fig:fig.2}
\end{figure}

The plots also show that $\Lambda$ production and annihilation through
inelastic re-scattering stops after a certain time (chemical freezeout).
It turns out, though, that this is not due to our model reaching strangeness
chemical equilibrium but rather based on the rapid expansion and the
resulting drop in particle density. After around 30 fm/c the particle
density is simply too small to support further inelastic scatterings which
produce strange baryons.
To prove that our model reaches equilibrium for the case of a non-expanding
system by properly envoking detailed balance in our cross sections, we chose 
a fixed maximum
radius of 10 fm and plotted the $\Lambda$ production and annihilation channels
as a function of time. From Fig.3 it is apparent that equilibrium is
reached, the total $\Lambda$ yield becomes constant, whereas the production 
and annihilation
yields continue to rise at long times. Thus chemical equilibration is possible
if the expansion velocity is sufficiently small. Our calculations agree
with other thermal model calculations, though, in predicting that the Pb+Pb 
system at CERN SPS energies does not reach chemical equilibrium before 
freeze-out.
Based on Fig.1 we can conclude that all particle ratios that include
the $\Lambda$ yield will
undergo significant changes during rescattering, a fact which is incompatible 
with the notion of sudden hadronization. 
If we force sudden hadronization in our code by simply turning off all
inelastic rescattering modes, our model calculations lead to a much 
smaller HBT radius in both the $\pi\pi$ and KK channels compared to the 
actually measured radii.

\medskip

\begin{figure}[hbtl]
\hspace{1.5in}
\psfig{figure=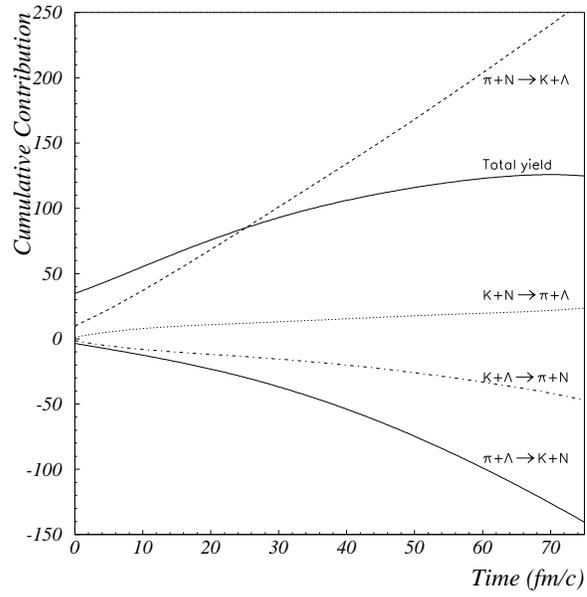,height=3.in}
\caption{Detailed balance in a system of fixed maximum radius (r=10 fm)}
\label{fig:fig.3}
\end{figure}

\medskip

\begin{figure}[hbtl]
\hspace{1.5in}
\psfig{figure=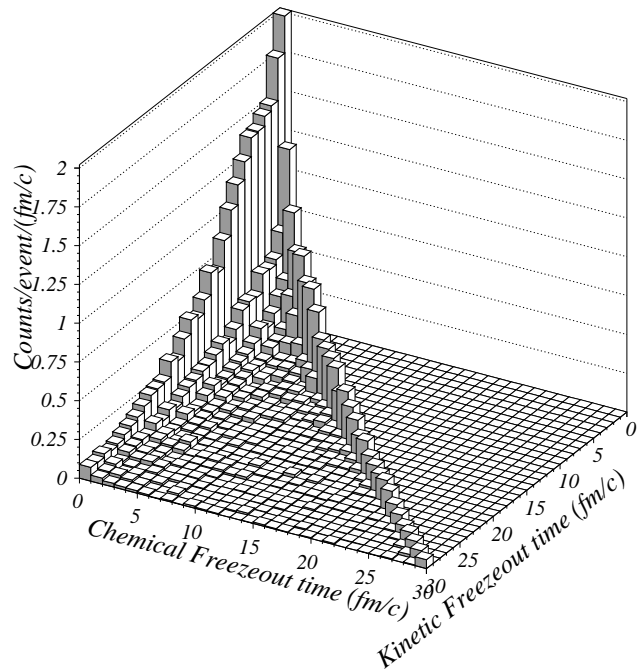,bbllx=100pt,bblly=100pt,bburx=520pt,bbury=645pt,height=3.5in}
\caption{Relation between chemical freezeout time and kinematic freezeout
time for measurable $\Lambda$ particles}
\label{fig:fig.4}
\end{figure}

\medskip

\clearpage

Fig.4 shows the relation between kinetic and chemical freezeout time
for every $\Lambda$ measurable in the Pb+Pb system after kinetic freezeout. 
It is apparent that only about half of the emitted $\Lambda$'s show a 
chemical freezeout time consistent with the hadronization time, which is 
required for sudden hadronization (t$_{chem}$=0,t$_{kin}$=any). 
Many $\Lambda$'s (about 25\%) are actually produced in their final rescattering 
step (t$_{chem}$=t$_{kin}$).

A comparison of the one-dimensional freeze-out time distributions for
the measured $\Lambda$'s in Fig.5 shows that the chemical freeze-out
occurs fast but it is not 'sudden'. 90\% of the $\Lambda$'s are chemically
frozen out after 5 fm/c, but of those about 40\% freeze-out after 
hadronization. 
The according kinetic freeze-out times are shown in Fig.6. Fig.6a shows the
difference in kinetic freeze-out time between the direct and the produced
$\Lambda$ component. The production after rescattering significantly
enhances the average freeze-out time and alters the shape of the
distribution. But the $\Lambda$ freeze-out is still peaked significantly
earlier than the pion and proton freeze-out in our model \cite{humanic2}
as seen in Fig.6b.
The main reason is that the
inelastic and elastic cross sections for $\Lambda$ induced reactions are
much smaller than the proton or pion interaction cross sections. This
trend will continue for multi-strange baryons and it should lead to a very
early decoupling of the $\Omega$ from the fireball \cite{xu}. For 
multi-strange baryons the notion of sudden hadronization seems thus more 
sensible.

\medskip

\begin{figure}[hbtl]
\begin{center}
\mbox{
\epsfig{file=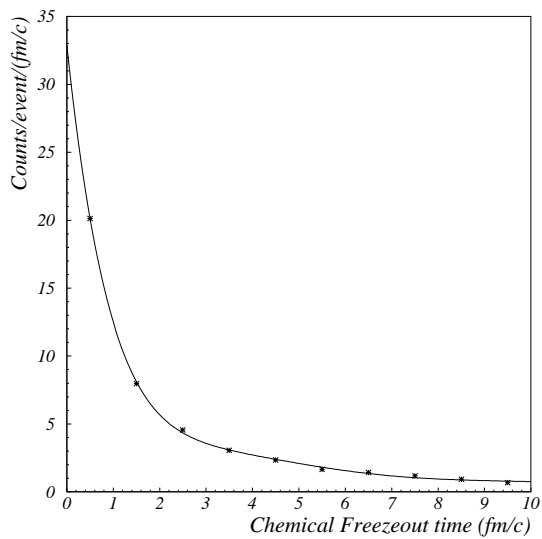,bbllx=50pt,bblly=155pt,bburx=520pt,bbury=645pt,width=6.5cm}
}
\end{center}
\caption{Chemical freezeout time for measurable $\Lambda$ particles.}
\end{figure}
\begin{figure}[hbtl]
\begin{center}
\mbox{
\epsfig{file=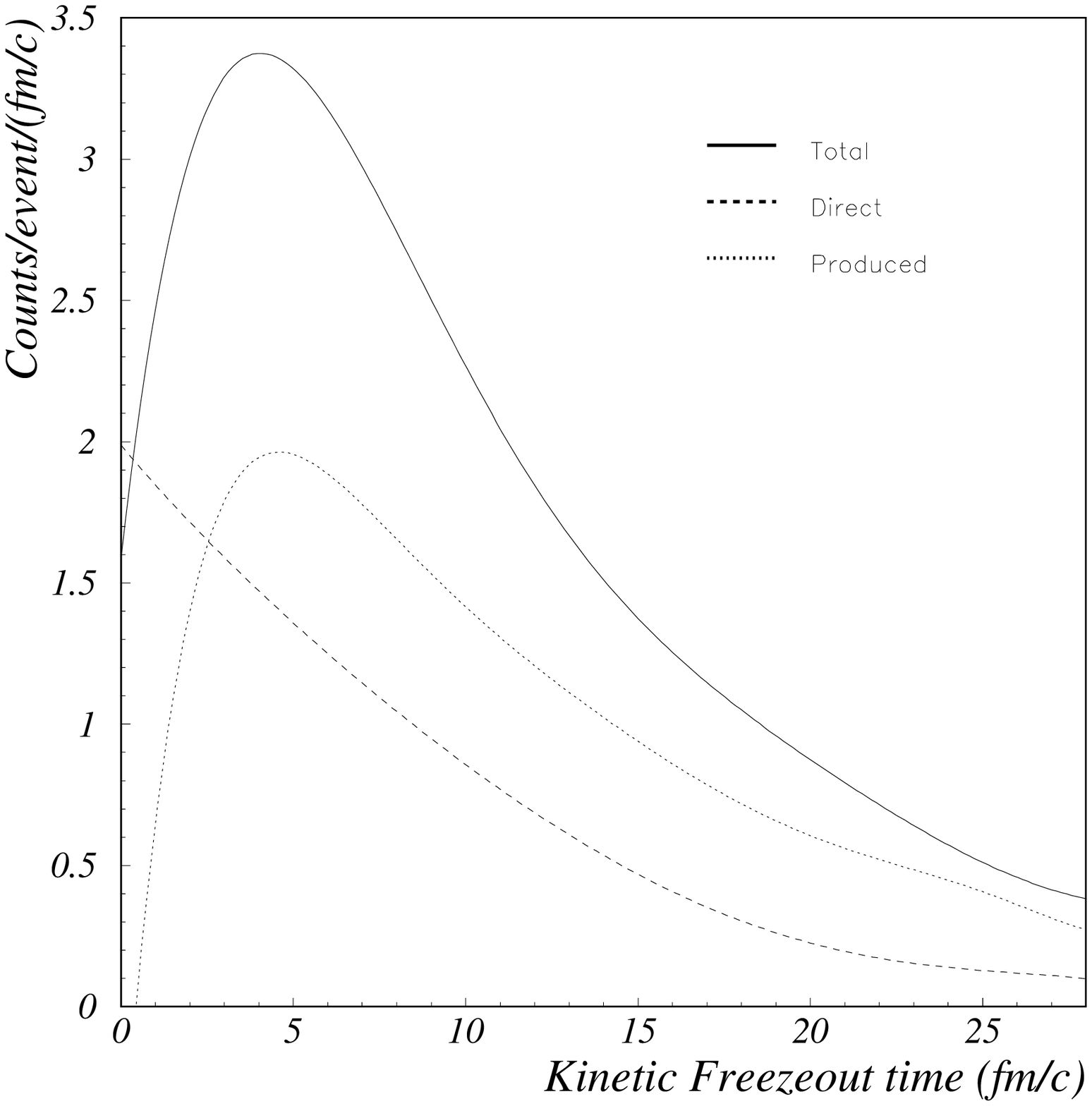,bbllx=50pt,bblly=155pt,bburx=520pt,bbury=645pt,width=6.5cm}
}
\mbox{
\epsfig{file=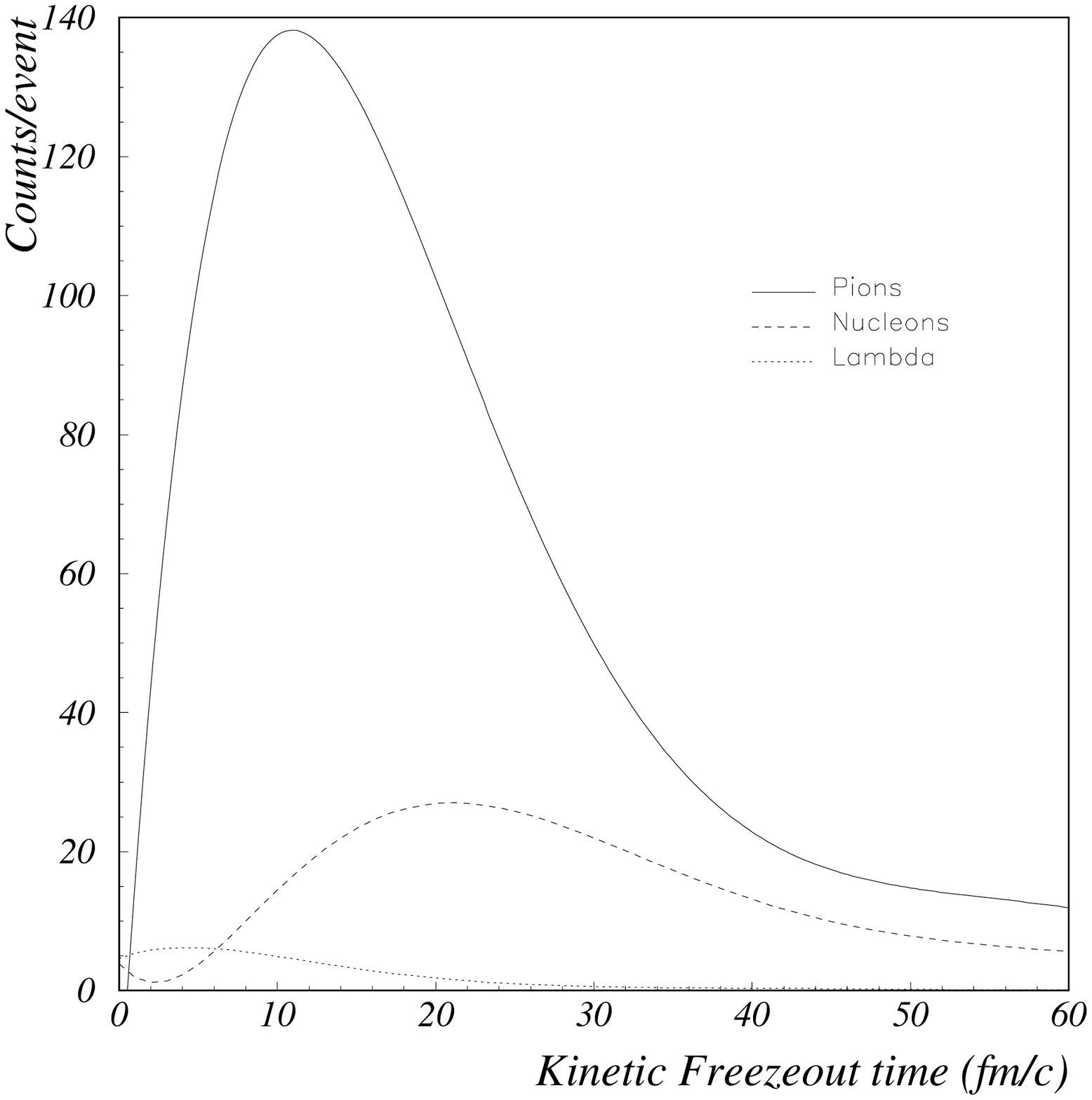,bbllx=50pt,bblly=155pt,bburx=520pt,bbury=645pt,width=6.5cm}
}
\end{center}
\caption{a.) Kinetic freezeout time for the different components of the
$\Lambda$ spectrum, b.) Comparison between the kinetic freezeout time of
measurable $\Lambda$'s , $\pi$'s and protons}
\end{figure}

\medskip

Fig.7 shows the transverse mass spectrum of the frozen out $\Lambda$'s. 
The model is in perfect agreement with the data measured by NA49 as it is
for the pions and nucleons.  

\medskip

\begin{figure}[hbtl]
\hspace{1.5in}
\psfig{figure=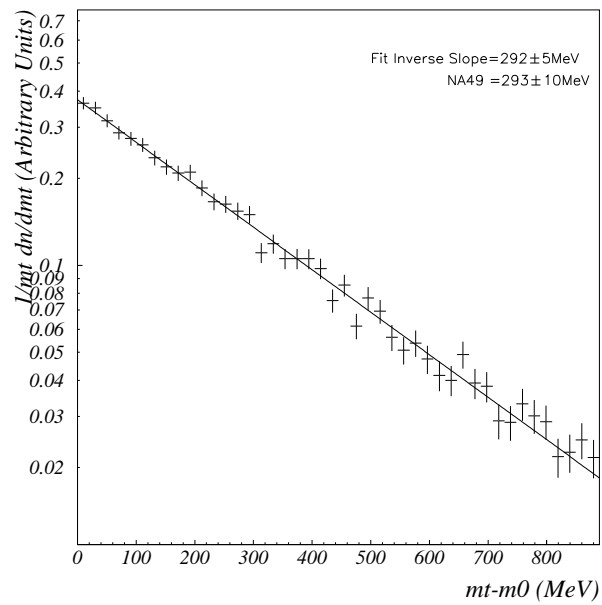,bbllx=50pt,bblly=155pt,bburx=520pt,bbury=645pt,height=3.in}
\caption{Transverse mass spectrum for $\Lambda$ from our calculations. 
The extracted slope parameter is seen to agree with data (NA49).}
\label{fig:fig.7}
\end{figure}

\clearpage

\bigskip
\medskip
\noindent{\bf Conclusions}
\medskip

Based on our calculations we conclude that neither the $\Lambda$ yield 
nor any particle ratio including $\Lambda$'s are suitable thermometers 
for the hadronization 
temperature. These measurements may still be indirect QGP indicators in any 
transport code description, though, simply through determination of the 
required initial hadronization conditions (initial temperature and energy 
density). This is a departure from the original wisdom that only leptonic
probes can act as real indicators of the initial conditions. 

We have shown that the $\Lambda$ abundance is strongly affected 
by final state interactions and is not reproducable under the 
condition of sudden hadronization, if the model attempts to describe
all hadronic measurements in particular in the heaviest SPS systems. 
The probability of inelastic scattering is sufficiently large that even during 
the rather short time lapse between hadronization and thermal freezeout the 
yields change considerably. Thus, conclusions drawn from the strange baryon 
ratios at CERN concerning a QGP phase transition are probably an 
oversimplification regarding the actual interactions inside the fireball.

Our simulations agree with other thermal models that the chemical properties
are frozen well before the kinetic spectra but we also demonstrate that 
hadronization and chemical freezeout mostly do not coincide for strange
baryons produced in Pb+Pb. Thus the  hadronization should occur above 
T=180 MeV which was deduced from the ratios as the temperature for chemical 
freezeout. Indeed, in our calculation the system hadronizes at 213 MeV, well 
above the critical temperature. 
It is also evident that both, the $\Lambda$ chemical and thermal freeze-out,
occur much earlier than the proton and pion freeze-out. 
Recent detailed simulations of multi-strange baryon ratios and spectra, seem
to indicate that this early decoupling follows a trend as a function of
the strangeness content. This effect can be attributed to 
differences in the rescattering probabilities \cite{dumitru}.
In particular the apparent lack of flow in the $\Omega$ transverse momentum 
spectrum could be correlated with the very high $\Omega$ particle ratios. 
Both point at a uniquely different production mechanism for the $\Omega$. 
In Dumitru's calculations the $\Omega$'s undergo on
average two collisions after hadronization, whereas the number increases
to 3.5 and 5 for $\Xi$ and $\Lambda$, respectively \cite{dumitru}. 
This early decoupling of the $\Omega$ led van Hecke et al. 
to postulate that this is the reason for the lower emission temperatures 
for $\Omega$'s as measured at the SPS \cite{xu}. Our calculation indicates
that even the $\Lambda$ thermally freezes out after a few fm/c,
due the rather small elastic scattering cross section during 
the rescattering phase. It is our goal to extend our transport code to 
include the multiply strange baryons to determine more quantitatively the
dynamic differences between strange and multistrange baryons.

The complete measurement of strange particle production at RHIC, from the 
Kaon to the $\Omega$, will be an important exercise. Our SPS study shows that 
final state interactions have to taken into account in order to properly
describe singly strange particle production. Thus, Kaon and $\Lambda$ 
measurements alone might not lead to conclusive proof of the formation of a 
QGP. However plans for detailed measurements of strangeness production up to 
and including the $\Omega$ are well underway at RHIC and we are looking 
forward to this exciting new era in relativistic heavy ion physics.

\end{document}